# Tunable Wettability of Polymer Films by Partial Engulfment of Nanoparticles

Weiteng Guo[1], Chongnan Ye[1], Gert H. tenBrink[1], Katja Loos[1], Vitaly B. Svetovoy[2], and George Palasantzas[1],*

[1]Zernike Institute forAdvanced Materials, University of Groningen, Nijenborgh 4, 9747 AG Groningen, The Netherlands

[2]A. N. Frumkin Institute of Physical Chemistry and Electrochemistry, Russian Academy of Sciences, Leninsky prospect 31 bld. 4, 119071 Moscow, Russia

## Abstract

A series of poly(methyl methacrylate) (PMMA) surfaces decorated by Cu nanoparticles (NP) with gradually varied morphology were prepared by high-pressure $CO_2$ treatment at various time spans. Combining the characterizations of transmission electron microscopy (TEM) and atomic force microscopy (AFM), an accurate 3-dimensional view of the morphology of the surfaces was presented. Subsequently, the wettability of the surfaces decreases near-linearly with the increase of the apparent height of the decorating NPs in both static (static contact angle) and dynamic (contact angle hysteresis) aspects. The observed tendency contradicts to the Wenzel or Cassie-Baxter model and is explained by the contribution of nanomeniscus formed between the decorating NP and the flat substrate. The capillary pressure from this meniscus is negative and results in the increase of the contact angle with the apparent height ($H_N$) of the Cu NPs decorating the PMMA surface. In addition, the effect of the coverage ($C_N$) by NPs on the wettability can be explained on the same basis. Our experiment demonstrates the important influence of the nanomeniscus on the wettability, which is usually not taken into account. The results in this work provide a comprehensive understanding of how nanostructure affects the wettability of the decorated surfaces and shed light on how to obtain certain wettability through nanostructuring of the surface morphology.

___

*Corresponding author: g.palasantzas@rug.nl

# I. INTRODUCTION

Nowadays relentless efforts have been devoted to creating different types of roughness to enable control of the wettability of surfaces. With high static contact angle (SCA) and low contact angle hysteresis (CAH), the lotus leaf [1] inspired various applications such as, for example, self-cleaning surfaces [2] and low-friction surfaces for fluid flow [3]. Unlike the lotus leaf, some rose petals [4], scallions, and garlic exhibit superhydrophobicity with high CAH [4–6], inspiring applications in fields such as droplet transportation [7] and energy harvesting [8].Although the wetting behavior of lotus leaf-like surfaces and rose petal-like surfaces differ a lot with each other, an important path to realizing the certain wettability is building up the microstructure of the surfaces. Therefore, understanding the relationship between wettability and detailed surface morphology is important for designing surfaces with certain wettability for various applications.

Many efforts have been paid for understanding the describing roughness and wettability, and Wenzel [9] and Cassie-Baxter [10] models are two most widely used ones. In the Wenzel theory, the testing droplet wets the cavities of the surface, enlarging the interaction area with the liquid droplet. And the apparent SCA $\theta_W$ in terms of the Wenzel model is given by the expression $cos\theta_W = rcos\theta_Y$ [9], where $r = A_r/A_p$ is the roughness factor with $A_r$ the actual rough surface area and $A_p$ the projected surface area on the average surface plane. However, in several cases, *e.g.*, the lotus effect, [11–13] there are air pockets trapped between the solid surface and the testing droplet. This case is described by the Cassie-Baxter model, where the apparent SCA $\theta_{CB}$ is given by the expression $cos\theta_{CB} = fcos\theta_y + (f - 1)$, [10] where $f$ is the fraction of the solid surface area wetted by the liquid.

Although the two well-known models have enabled the explanation of the wettability of surfaces with various types of morphology, several other studies have demonstrated complex cases of surface wettability, where the Wenzel and Cassie-Baxter models could not explain the experimental data. Taking the air-liquid interface on a solid as one-dimensional system, Pease [14] emphasized that the SCA is the result of an equilibrium position that the three-phase contact line (TPL) could reach. This perspective was supported by experiments, where the pinning effect of the contact line was observed on micro-structured [15] and nano-structured surfaces [16], and extensively discussed and summarized by Gao and McCarthy [17]. These works illustrated that the reason why the Wenzel model failed to predict the SCA of micro-structured surfaces was that a thermodynamic description could not account for contact line pinning. Therefore, the local equilibrium of the TPL is critical to describe the relationship of the wettability and geometrical features (roughness), inspiring more studies to explore the effects of macro-sized or nano-sized decorations on wettability. Using micro-sized square pillars with various sizes and concentrations, Forsberg *et al.* [18] revealed the microscopic details of the contact line pinning behavior, and bridge the relationship between the wettability and the varied micro-sized decorations. Furthermore, the contact line pinning behavior could be analyzed to a nanometer scale, [16,19] where the mechanism of pinning at a nanometric scale might be attributed to the nanomeniscus formed between the substrate and the nanoscale decoration [20]. Besides, simulations of the contact line pinning effect on nanoscale textured surfaces [21] and the pinning effect of a single nanoparticle [22] were conducted to give a deep understanding of the relationship between wettability and nanostructure. However, experimental investigations taking into account nanoscale defects are still limited, because such studies are faced with the difficulty of obtaining and characterizing rough surfaces having a controlled morphology of the nanostructure. Therefore, the goal of the present work is to create a series of surfaces with gradually varied nanostructure to understand the influence of nanometric decorations on the wetting properties of a solid surface.

In this framework, nanoparticles (NPs) provide an ideal option to create nanostructured surfaces and achieve certain wettability, *e.g.*, $TiO_2$ NP painting [23]. or even more directly, *e.g.*, Cu NP deposition by a high-pressure magnetron sputtering system where the production and decoration process can be accomplished in one step [24]. Moreover, the high-pressure magnetron sputtering can offer homogeneously distributed Cu NPs onto flat substrates without introducing additional chemical ligand [24,25], which made it an ideal candidate to provide nanostructured decoration for wetting research. Besides, Teichroeb *et al*. [26] studied the embedding of gold nanoparticles into the surface of polystyrene (PS) to probe the viscoelasticity of polymer surfaces. Subsequently, Tan *et al.* [27] reported the controlled thermally assisted particle embedding of surface deposited silica NPs at the surface of poly(methyl methacrylate) (PMMA) polymer films. Particle embedding was controlled by varying the temperature and time of the thermal treatment, and similar results were obtained for surface-modified silica NPs in PMMA and poly(methyl methacrylate-co-methacrylic acid) films [28]. As an alternative to thermal annealing, Lee *et al*. [29] and Yang *et al.* [30] reported the embedding of gold NPs in PS films via $CO_2$ saturation of the polymer substrate at relatively low temperatures, because $CO_2$ saturated PS surface exhibits an increased polymer mobility, allowing the particles to sink into the surface. Furthermore, Liu *et al.* [31] gave extensive experimental results and theoretical analysis of the high-pressure $CO_2$ assisted engulfment of NPs on PMMA, proving the high-pressure $CO_2$ a reliable method to conduct NP engulfment on polymer films. Therefore, we have gathered ideal candidates of nanostructured decoration (magnetron sputtered Cu NPs) and substrate (PMMA) to design nanostructured surfaces for wetting research.

Here we used the high-pressure $CO_2$ assisted NP engulfment technique, and designed a series of samples with Cu NPs decorated PMMA surfaces with gradually varied apparent height of NPs by controlling the time span of high-pressure $CO_2$ treatment. Taking advantage of the superior lateral resolution of transmission electron microscopy (TEM) and vertical resolution of atomic force microscopy (AFM), an accurate 3-dimensional view of the morphology of the surfaces was presented. Subsequently, wetting measurements of SCA as well as advancing and receding contact angle (ACA/RCA) were conducted on the surfaces with controlled nanoscale morphology. Finally, a model involving the wettability contribution of the nanomeniscus formed between the decorating NP and the flat substrate was proposed to describe the relationship between the wettability and apparent height ($H_N$) of the Cu NP decorating PMMA surfaces.

## II. EXPERIMENTAL RESULTS AND DISCUSSION

Here we will present our main experimental results and discussion about them, while details about the preparation methods and characterization techniques are given in the Appendix.

### (a) Wettability of PMMA films

The PMMA films were made using the same method as in the work of Liu *et al*. [31] The thickness of the PMMA films was measured to be about 50 μm, [31] which is sufficiently thick to be taken as a bulk in the wetting study in this work. Additionally, the PMMA films on silica wafers were very smooth (see Fig. 1(a)), referring them as very suitable substrates for the current wetting research. In addition, the wetting measurements of the different PMMA films showed stable wettability (with SCA 81.5±0.5° in Fig. 1(b), and ACA 83°, RCA 70.5±0.5° for dynamic wettability). Besides the stable wettability, also the morphology of the PMMA films did not show any significant variation. On behalf of checking the effect of the high-pressure $CO_2$ treatment to the morphology and wettability of PMMA films, we put an extra PMMA film together with the NPs/PMMA samples during each high-pressure $CO_2$ treatment. As a result, there was no evident variation among the wettability and the height distributions (see Figs. 1(c)-(f)) of the PMMA surfaces after high-pressure $CO_2$ treatments with various time spans.

### (b) Effect of Cu NPs on wettability of PMMA films

In our previous study [20], a hypothesis has been formulated that the increase of the SCA of a surface could be generated by the nanomeniscus, whose existence has been proved by experiments [32–34] and theoretical analysis [35–37], formed between the substrate and the decorating NPs. Subsequently, it is reasonable to assume that the effect would become weaker when the NPs are partially submerged into the substrate, and the untreated samples have to demonstrate the highest SCA. Therefore, the first step to check the hypothesis is to allow the Cu NPs to induce a difference in the wettability of the PMMA and then measure the variation of the wettability after high-pressure $CO_2$ treatments. Knowing that the surfaces with higher NP coverage $C_N$ (the ratio of the area covered with NPs with respect to the overall area as determined by the TEM images, which is the same definition used in our previous works [20,24]) tend to show large SCA [24], we started from the preparation of high-$C_N$ Cu NPs/PMMA samples. The morphology as well as the SCA of the Cu NPs decorated PMMA samples are shown in Figs. 2(a), (b). The $C_N$ was measured to be about 64%, and the SCA of the PMMA film has experienced a significant increase from 81.5±0.5°(see Fig. 1(a)) to 122±1° (see Fig. 2(b)) after the deposition of Cu NPs. However, the in-plane resolution (see Supplemental Material for details [42]) of the AFM images is not good enough to distinguish two NPs close to each other [38] because of the tip shape effect [39,40]. When the $C_N$ is too high, *i.e.*, 64% in this case, the individual NPs would be too close to each other for the AFM tip to distinguish them. As a result, the AFM image would appear to be totally "covered" by NPs, and apparent heights of the individual NPs are hardly accessible for so high $C_N$ sample, though the AFM vertical resolution is outstanding [41]. Therefore, lower $C_N$ PMMA films are necessary for obtaining the apparent heights of the individual decorating NPs.

### (c) Partial NP engulfment in PMMA

Since the samples with high $C_N$ were not suitable for height measurement of the individual decorating NPs, low-$C_N$ Cu NP/PMMA samples were prepared, offering sufficient flexibility for the size analysis of the Cu NPs. Since an NP could completely sink into PMMA with high-pressure $CO_2$ treatment when its diameter is below 12 nm [31], the deposition parameters of the sputtering system for Cu NPs were optimized to obtain NPs with a diameter of approximately 12 nm (the method for controlling the diameter of Cu NPs was discussed in our previous work [25])so that one can achieve various degrees of engulfment in the PMMA film. The coverage of the samples was measured to be about 12% in Fig. 2(c), which offers NPs with good monodispersity for dimensional measurements. In fact, image analysis with the Image-Pro Plus v6.0 software yielded the distribution of the diameter ($D_N$) of the NPs as it is shown in Fig. 2(d).The median value of $D_N$ was 11.7 nm by fitting with a Gaussian function (more than 220 individual isolated NPs were measured), while the standard deviation (*SD*) was 2.2 nm, indicating a good fit by the Gaussian model as well.

The SCAs of the low (12%)-$C_N$ samples were measured to be approximately 81°, showing no noticeable difference in wettability from bare PMMA films (81.5±0.5°, Fig. 1). However, the good NP monodispersity makes the low $C_N$ samples suitable candidates for AFM analysis. The apparent heights ($H_N$) of the individual NPs(see Supplemental Material [42] for details) on the low (12%)-$C_N$ samples before high-pressure $CO_2$ treatment were derived from Fig. 2(e) and summarized in Fig. 2(f). The median of $H_N$ (for meaningful statistics the value of $H_N$ for 270 individual isolated NPs were measured) was evaluated as 12.1 nm by fitting with the Gaussian function, while the obtained *SD* was 2.1 nm indicating that the fit of the data by the Gaussian function is good. Moreover, the median of $H_N$ (12.1 nm) agrees well with the median of $D_N$ (11.7 nm shown in Fig. 2(d)) which was measured in the TEM image (Fig. 2(c)), indicating the absence of NP engulfment before high-pressure $CO_2$ treatment. In addition, the agreement between the TEM and AFM studies represent a consistency check of the lateral/vertical resolution of the TEM/AFM images. The latter could allow the reconstruction of an accurate 3-dimensional view of the samples. Additionally, no obvious deposition-caused damage was observed when comparing Fig. 1(a) and the NPs-uncovered areas in Fig. 2(e), indicating the magnetron sputtering system used in this work would not do damage to the PMMA films.

Subsequently, high-pressure $CO_2$ treatments of 30 min and 90 min were conducted to the low-$C_N$ samples, respectively. The corresponding AFM images are shown in Fig. 2(g) and Fig. 2(i), respectively. Then the height $H_N$ of 270 isolated individual NPs were measured, and the results were summarized in Figs. 2(h) and 2 (j). The median of $H_N$ of the 30-min $CO_2$ treated sample was evaluated to be 5.9 nm by the Gaussian fitting with an *SD* of 1.5 nm indicating a good fit.

However, for the 90-min $CO_2$ treated sample, the median of $H_N$ was evaluated to be 3.0 nm with a large *SD* of 2.8 nm. The latter can be understood by the observed asymmetry of the distribution for $H_N$ in Fig. 2(j), which may be caused by the uncertainty of measuring the NPs with $H_N$ lower than 2 nm in the AFM image. Comparing the results in Figs. 2(e)-(j), it can be concluded that the Cu NPs are in metastable mechanical state and gradually sink into PMMA films during the high-pressure $CO_2$ treatment, where the NPs will undergo more engulfment with more extended high-pressure $CO_2$ treatment [31]. Additionally, the number density as well as the distribution of decorating NPs did not vary a lot in the AFM images shown in Figs. 2(e), (g), (i), indicating that all the individual NPs sank uniformly during the high-pressure $CO_2$ treatments. Therefore, the high-pressure $CO_2$ treatment used in this work is an ideal method for controlling the $H_N$ of NPs on PMMA films and creating a series of Cu NPs/PMMA samples with comparable morphologies.

### (d) Wettability *vs*. NP engulfment into PMMA

The Cu NPs/PMMA samples with the "intermediate"-$C_N$ (37%) were studied extensively. The morphology studies of the NP engulfment into PMMA and the wettability studies of the relevant samples are shown in Figs.3 and 4, respectively. Besides the AFM images in Figs.3(a), (c), (e), (g), the height $H_N$ of the isolated NPs were also measured in two additional AFM images (see Supplemental Material [42] for details) derived from each sample to improve statistics. In fact, the $H_N$ of more than 120 individual isolated particles were analyzed for each $H_N$ distribution and shown in Figs.3(b), (d), (f), (h). The distribution of $H_N$ in the "intermediate" (37%)-$C_N$ sample before any high-pressure $CO_2$ treatment was obtained from Fig. 3(a) and shown in Fig. 3(b). The median of $H_N$ was evaluated to be 14.0 nm (by fitting with a Gaussian function with *SD*=1.4 nm). Additionally, the median of $D_N$ was evaluated to be 13.2 nm (see Supplemental Material [42] for details). Similar to the agreement between the results in Figs. 2(f) and 2(d), the median of $H_N$ (14.0 nm) agrees well with the median of $D_N$ (13.2 nm) indicating the absence of any NP engulfment before the high-pressure $CO_2$ treatment, and the satisfactory accuracy of the lateral/vertical resolution of the TEM/AFM images.

Subsequently, three "intermediate" (37%)-$C_N$ samples were treated separately by high-pressure $CO_2$ for 5 min, 30 min, and 90 min, respectively. The AFM images and distributions of $H_N$ are shown in Fig. 3. All the distributions of $H_N$ were fitted by the Gaussian function, and the medians of $H_N$ for the 5-min, 30-min, and 90-min $CO_2$ treated samples were 10.1 nm ($SD = 1.3\ nm$), 5.8 nm ($SD = 2.6\ nm$), and 4.7 nm ($SD = 3.6\ nm$), respectively. The results in Fig. 3 are similar to the ones in Figs. 2(e)-(j), reconfirming that the high-pressure $CO_2$ treatment used in this work is a reliable way for NP engulfment into PMMA.

Subsequently, the wettability of the "intermediate" (37%)-$C_N$ samples was measured and summarized in Table 1 and Figs. 4(a), (b). The square (blue) spots in Fig. 4(a) show the evolution of the SCAs starting from 81.5±0.5° (bare PMMA film) to 99° (NPs/PMMA without high-pressure $CO_2$ treatment). The SCA appears to increase almost linearly as a function of medians of $H_N$, where $\chi^2 = 0.99$ with linear regression shown in Fig. 4(a) ($\chi^2$ is the coefficient of determination defined as $\chi^2 = 1 - SS_{RES}/SS_{TOT}$). Considering the fact that the high-pressure $CO_2$ treatment changed neither the morphology nor the wettability of the PMMA films (see Fig. 1), the SCA of the Cu NPs/PMMA samples can only be associated to the variation of $H_N$ of the decorating NPs.

Furthermore, the ACAs and RCAs of the samples were measured to explore the dynamic wettability of the samples, and the results are summarized in Fig. 4(b). In general, the results of ACA show a similar tendency to the results of SCA. Specifically, the difference between bare PMMA film and the 90 min high-pressure $CO_2$ treated sample for the ACA (~11°) was more significant than the one in SCA (~5.5°), indicating that the decorating NPs might play a more important role in dynamic wettability as compared with the static one. Moreover, this phenomenon was more dramatic for the results of RCA. In general, the RCAs of the samples showed an inverse tendency as compared with that of ACAs and SCAs, while the "rate" of variation is higher compared with the ones of the ACA and SCA.

According to the definition of the CAH $\theta_H = \theta_A - \theta_R$ (with $\theta_A$ the ACA and $\theta_R$ the RCA) [43], the $\theta_H$ is marked as the dashed areas in the (green) bars of Fig. 4(b), while it is shown as the circular (red) spots in Fig. 4(a). Similar to the dependence of SCA from $H_N$, the relationship between the CAH and $H_N$ also followed a near-linear behavior with $\chi^2 = 0.97$ by linear regression. However, the slope of the profile of CAH-$H_N$ is larger than the one of SCA-$H_N$, suggesting again that the dynamic wettability is more "sensitive" to decorating NPs than the static one.

| $H_N$ (nm) | SCA(°) | ACA(°) | RCA(°) | CAH(°) |
|---|---|---|---|---|
| 0 | 81.5±0.5 | 83 | 70.5±0.5 | 12.5±0.5 |
| 4.7 | 87±1 | 94±2 | 67 | 27±2 |
| 5.8 | 90 | 98±1 | 63 | 35±1 |
| 10.1 | 93.5±0.5 | 101±1 | 56.5±2.5 | 44.5±3.5 |
| 14.0 | 99 | 105.5±2.5 | 36.5±1.5 | 69±4 |
| $C_N$ (%) | | | | |
| 0 | 81.5±0.5 | 83 | 70.5±0.5 | 12.5±0.5 |
| 12 | 81 | 89±1 | 65.5±1.5 | 23.5±2.5 |
| 27 | 85±1 | 98±1 | 59.5±1.5 | 32.5±2.5 |
| 37 | 99 | 101±1 | 36.5±1.5 | 69±4 |
| 46 | 107±2 | 105.5±2.5 | 29±1 | 82±2 |
| 64 | 122±1 | 124.5±0.5 | 24.5±1.5 | 100±2 |

**Table 1** Wettability data of the "intermediate" (37%)-NP coverage Cu NPs/PMMA samples with various $H_N$, and the Cu NPs/PMMA samples with various $C_N$ prior to high-pressure $CO_2$ treatments.

In addition, we have also summarized the SCAs derived from the Cu NPs/PMMA samples with various $C_N$ values (see Table 1), as it is shown by the square (blue) spots in Fig. 4(c). The profile starts from the PMMA film with an SCA of 81.5±0.5°, undergoing no notable change before the SCA of 37%-$C_N$ sample (99°). Subsequently, the SCA increased in an almost linear manner. Aiming to provide more supporting information by Fig. 4(d), the ACAs and RCAs were also measured from the samples and summarized in Table 1. In detail, the difference between bare PMMA film and the 12%-$C_N$ sample in ACA (~6°) was more significant than the one in SCA (~-0.5°), reconfirming the conclusion derived from the results in Figs. 4(a), (b) that the decorating NPs likely play more important role in the dynamic than the static wettability.

Again, the RCAs of the samples showed an inverse behavior compared with the one of the ACAs or SCAs. Notably, the RCA displayed a sudden drop between the 27%-$C_N$ and the 37%-$C_N$ sample with a difference of 23°. Also, the results of CAH are marked as the dashed areas in Fig. 4(d). In detail, the gap between the 27%-$C_N$ sample and the 37%-$C_N$ sample in CAH (~36.5°) is much larger than the one in SCA (14°). On the one hand, the dynamic wettability analysis reconfirmed that the 37%-$C_N$ was located near the critical point of the NP coverage, where the wettability changed strongly enough to indicate that this sample was a good candidate for the subsequent wettability-NP engulfment investigation. On the other hand, the comparison between the results for CAH and SCA supported the suggestion obtained from Figs. 4(a) and (b) that the dynamic wettability is more "sensitive" to the decorating NPs than the static one.

## III. NANOMENISCUS EFFECT ON WETTABILITY

Neither of the two widely-used Wenzel [9] and Cassie-Baxter [10] models is suitable for analyzing the wettability of the Cu NP decorated PMMA surfaces. On the one hand, the materials used in this work are relatively hydrophilic with similar contact angles (81.5±0.5° for PMMA as the substrate material and 79° for Cu [20] as the NP material), which makes impossible to generate hydrophobicity within the Wenzel model. On the other hand, the Cu NPs decorated PMMA surfaces were not in the Cassie-Baxter state, because the measured large contact angle hysteresis (see Figs. 4(a), (c)) indicates strong adhesion force between the water meniscus and the tested surface. A similar situation for a surface decorated by NP was discussed previously [20]. It was proposed that the paradox can be resolved by taking into account the formation of a nanomeniscus when liquid wets a spherical particle on a flat surface. This concave nanomeniscus shown schematically in Fig. 5(a) will give a negative contribution to SCA, increasing the effective contact angle of the decorated surface.

Let $\theta_S$ be the SCA of the substrate and $\theta_N$ is that for the NP material. In the Wenzel model, the effective SCA can be calculated according to the relation

$$\cos\theta_W = f_S \cos\theta_S + f_N \cos\theta_N, \tag{1}$$

where $f_S$ and $f_N$ are the ratios of the true area of the solid surface to the apparent area for the substrate and NP, respectively. For a sphere of radius $R$ submerged to the substrate to a depth $H$ these ratios are

$$f_S = 1 - C_N \frac{H(2R-H)}{R^2}, \quad f_N = C_N \frac{2(2R-H)}{R}, \tag{2}$$

where $C_N$ is the surface coverage by the NPs(more details for the derivation of Eq. (2) is seen in Supplemental Material [42]). The contact angle on the flat substrate is defined by the Derjaguin equation [44,45]

$$\cos\theta_S = 1 + \frac{1}{\gamma}\int_{h_S}^{\infty} \Pi_S(h)dh + \frac{h_S}{\gamma}\Pi(h_S). \tag{3}$$

In Eq.(3) $\gamma = \gamma_{lv}$ is the surface tension of the liquid, and $h_S$ is the equilibrium thickness of the wetting film and $\Pi_S(h)$ is the disjoining pressure on the substrate. For macroscopic liquid volumes the capillary pressure $|P_c| \sim \gamma/\mathcal{R}$ is small, where $\mathcal{R}$ is a macroscopic radius of curvature of the meniscus. In this case, the last term in Eq. (3) is small $h_S/\mathcal{R} \ll 1$ since $h_S$ is in the nanometer range. For this reason, we neglected the small term $h_S\Pi_S(h_S)/\gamma$ in Eq. (3) [46]. If the wetting film is thin in comparison with the size of NP $h \ll R$ and the particle is isolated, one could write a similar expression for $\cos\theta_N$ just changing $\Pi_S \to \Pi_N$ and $h_S \to h_N$. However, the nanomeniscus that is formed due to contact of the particle and substrate also gives a contribution to the contact angle and $\cos\theta_N$ can be presented as [20]

$$\cos\theta_N = 1 + \frac{1}{\gamma}\int_{h_N}^{\infty}\Pi_N(h)dh - \int_{h_2}^{\infty}\left(\frac{1}{r_1} - \frac{1}{r_2}\right)dh, \tag{4}$$

where $r_1$ is the negative radius of curvature (shown in Fig. 5(a)), and $r_2$ (not shown) is the positive radius of curvature in the orthogonal direction to the plane. These radii always have different signs and obey the condition $r_1 \ll r_2$. This condition guarantees that the contribution of the meniscus is always reducing $\cos\theta_N$. Following the procedure described in the work of Boinovich *et al*. [47], in principle, one can take into account the effects of the order $h/R$.

In addition, there were some clusters formed by several individual spherical NPs on the samples shown in Figs. S3 (a), (c), (e). Then the clusters could be taken as "lager NPs" with varied morphology. Subsequently, the shape of the nanomeniscus formed between the cluster and the substrate would change simultaneously. The curvature radius $r_1$ and $r_2$ in Eq. (4) would be lager compared with the situation of an individual NP. However, the increase of $r_1$ will be more significant than that of $r_2$, because the morphology of the clusters tends to be disc like. As a result, the $\cos\theta_N$ calculated by Eq. (4) would become larger, leading to a lower CA compared with the one of the situations we would apply in the following analysis where all the NPs would disperse perfectly.

Let us denote the last term in Eq. (4) as

$$\Delta(\cos\theta_N) = -\int_{h_2}^{\infty}\left(\frac{1}{r_1}-\frac{1}{r_2}\right)dh. \tag{5}$$

It is responsible for the transition to hydrophobicity, but its direct evaluation is not a simple task. This term cannot be expressed only via the disjoining pressures $\Pi_S(h)$ and $\Pi_N(h)$ in the liquid films on the substrate and on the particle, respectively, because near the apex of the meniscus, a significant contribution comes from the interaction between these films. Moreover, pressure in the liquid gets a nondiagonal tensorial structure that influences the mechanical equilibrium [48]. Equation of mechanical equilibrium in these conditions has been deduced [49], but practically it can be used only for a few simple problems. While we have no a reliable way to estimate $\Delta(\cos\theta_N)$ it is possible to extract some information from the experimental data.

The observed linear dependence of the SCA on the apparent height $2R - H$ suggests that $\Delta(\cos\theta_N)$ is a quadratic function of the parameter $x = H/R$. It is clear from the fact that according to Eqs. (1) and (2) $\cos\theta_W$ is quadratic in $x$. To cancel the quadratic term $\Delta(\cos\theta_N)$ also has to include the term proportional to $x^2$. The best linear fit of the data for $\cos\theta_W$ gives $\Delta(\cos\theta_N) = -(0.3955 - 0.0728x + 0.0006x^2)$. As obvious from Fig. 5 the curvature radius $r_1$ increases when the particle submerges deeper. When the particle sunk completely, this radius has to be infinite and $\Delta(\cos\theta_N) = 0$. The best fit gives $\Delta(\cos\theta_N) = -(0.3955 - 0.0728x + 0.0006x^2)$. Therefore, it is expected that $\Delta(\cos\theta_N)$ has to disappear with the apparent height at $x = 2$. This is not the case, and we have to include explicitly the zero at $x = 2$ in the function $\Delta(\cos\theta_N)$. The behavior of this function at $x \to 2$ follows from a simple analysis. For $x > 1$ the angle $\alpha$ (formed by the plane of the substrate and the tangent plane of the NP, which also passes through the intersection point of the substrate and the NP as it is shown in Fig. 5(b)) at the base of the submerged spherical particle is $\alpha = \frac{\pi}{2} - \sin^{-1}(x-1)$. In the limit $x \to 2$, it behaves as $\alpha \to \sqrt{2(2-x)}$. In this limit, the same behavior is expected for $\Delta(\cos\theta_N)$ that is why at arbitrary $x$ we are looking the function the form

$$\Delta(\cos\theta_N) = -(a_0 + a_1 x + a_2 x^2)\sqrt{1 - x/2} \quad (0 \leq x \leq 2). \tag{6}$$

The best fit of the SCA data in Fig. 4(a) gives $a_0 = 0.3955, a_1 = 0.0261, a_2 = 0.0195$. The function (6) is shown in Fig. 6(a). It decreases nearly linear with the apparent height, but at a height below 4 nm, it is going to zero fast. Fig. 6(b) shows the experimental data (red circles), the linear fit of the data $\theta_W = 81.70 + 1.228x$ (red dashed line), and the function that corresponds to Eq. (1) with $\Delta(\cos\theta_N)$ given by (6) (blue solid curve). One can see that the solid curve describes the data as well as the linear fit.

Additionally, the SCA also increases with the increase of $C_N$ (see Fig. 4(c)). Unlike the SCA-$H_N$ (see Fig. 4(a)) which shows a near-linear relationship, there is an obvious threshold of $C_N$ which "divide" the SCA-$C_N$ profile into two near-linear parts (see Fig. 4(c)). Therefore, more efforts should be paid for exploring this threshold and describing the SCA-$C_N$ relationship. Although the Wenzel model (1) does not predict the threshold, it is clear that it has to exist since one particle on a large area cannot influence the contact angle. Above the threshold $\cos\theta_W$ behaves linearly with the coverage. Modeling the threshold with a smooth transition, we look for the $C_N$ dependence in the form

$$\cos\theta_W = \cos\theta_S + \frac{a(x)}{2}\left(1 + \tanh\frac{C_N - C_N^0}{\Delta C_N}\right) C_N, \tag{7}$$

where $C_N^0$ is the position of the threshold and $\Delta C_N$ is its width. The amplitude $a(x)$ depending on the parameter $x$ is equal

$$a(x) = (2 - x)[-x\cos\theta_S + 2\cos\theta_N + 2\Delta(\cos\theta_N)]. \tag{8}$$

To compare this dependence with the data in Fig. 4(c) we take $x = 0$ that gives $a(0) = 4(\cos\theta_N - 0.3955)$. For this case the best fit of the data is shown in Fig. 4(c) by the red dashed line. It can be seen that such a dependence fails to describe the data at high coverage. One would expect that above the threshold, the coverage is not reduced exactly to a monolayer of NPs. If we allow the amplitude $a(0)$ to be augmented by a factor $\kappa$, then the data are well described by the dependence (7) as shown by the blue solid line. The parameters corresponding to the best fit are $\kappa = 1.278, C_N^0 = 0.3202, \Delta C_N = 0.0868$. They correspond to the average distance between particles at the threshold equal $3.1R$. It means that any additional particle will form the second layer that justifies the value of $\kappa$ larger than 1. In addition, the CAH data of the Cu NP decorated PMMA surfaces in Figs. 4(a), (c) showed similar tendency as the SCAs of the relevant samples, which agreed with the theoretical model in the aspect of dynamic wettability.

## IV. CONCLUSIONS

Using the technique of high-pressure $CO_2$ assisted NP engulfment, a series of Cu NP decorated PMMA surfaces with gradually varied nanostructured morphology were prepared by varying the time span of the high-pressure $CO_2$ treatment. Combining the characterization of transmission electron microscopy (TEM) and atomic force microscopy (AFM), an accurate 3-dimensional view of the morphology of the surfaces was presented. Subsequently, both static and dynamic wettability of the latter samples showed a near-linear tendency with the increase of apparent part of the decorating NPs. Finally, the relationship between the wettability and NP apparent height ($H_N$) of the Cu NP decorated PMMA surfaces was theoretically explained by evaluating the wettability contribution of the nanomeniscus formed between the decorating NP and the flat substrate. In addition, the effect of the coverage ($C_N$) of NPs on wettability was also checked compared with experimental results and theoretical analysis. In summary, our results provide a comprehensive understanding of how nanostructure affects the wettability of the decorated surfaces and shed light on obtaining certain wettability through structuring the surface morphology.

## Acknowledgements

We would like to thank Prof. Francesco Picchioni, and Mr. Marcel de Vries for their kindly offering the guidance and facilities for high-pressure $CO_2$ treatment. We would also like to acknowledge useful discussions with Prof. B. J. Kooi about the TEM characterizations. Moreover, we would like to acknowledge financial support from the China Scholarship Council (W.G.). Finally, we would like to acknowledge support from the Zernike Institute for Advanced Materials.

## APPENDIX: PREPARATION METHODS AND CHARACTERIZATION

### (a) Fabrication of PMMA substrates, Cu NPs deposition, and High-pressure $CO_2$ assisted NP engulfment

PMMA films were prepared by drop casting a PMMA-chloroform solution (0.1 g/mL) onto silica wafers (1cm × 1cm). The PMMA we used is a commercial product labeled with "Diakon LG156", and the properties of the polymer are available on the website [50], and we suppose this PMMA is replaceable by the PMMA with other brands or molecular weights. Afterwards, the substrates were dried in air for 24 h and then annealed at 135 °C for 12 h, and slowly cooled down to room temperature within the furnace. Additionally, the thickness of PMMA films on the silica using the same method was measured by Liu *et al*. [31] to be around 50 μm, indicating that the PMMA films were thick enough for this work because this thickness is much larger than the diameter of the deposited Cu NPs.

The Cu NPs were deposited in a modified *Mantis Nanogen* 50 system on the PMMA substrates, with the Cu NPs having a native surface oxide layer. A TEM grid with a continuous carbon supporting film was put together with the PMMA substrates for subsequent TEM observations. Then the Cu NPs were deposited simultaneously on PMMA substrates and the TEM grids in the magnetron sputtering system to ensure that the NP distribution on the Cu NP/PMMA samples and the TEM grid are the same. Moreover, the size and coverage of the Cu NPs were controlled by the settings in the *Mantis Nanogen* 50 system. More details can be seen in our previous work [25].

Finally, the NPs/PMMA samples were placed inside a high-pressure vessel. The setup temperature was 40 °C, and the pressure of $CO_2$ was 58 bar. Moreover, the time span for the $CO_2$ treatment was controlled to generate samples with various morphologies. After each $CO_2$ treatment, the valve connecting the high-pressure vessel and the open air would be slowly opened releasing the $CO_2$ into the open air.

### (b) TEM, AFM, and SEMcharacterization

The morphology of the as-deposited NPs was characterized using a JEOL 2010 at an acceleration voltage of 200 kV to record the bright-field TEM images of the Cu NPs on the TEM grids. Furthermore, the AFM images were obtained by a Bruker Multimode 8 AFM using tapping mode with a silicon cantilever (HQ:NSC15/No Al) having a resonance frequency of 325 kHz and a spring constant of 40 N/m. Moreover, the apparent heights of individual NPs were analyzed based on the obtained AFM images. The method for deriving the apparent height of an individual isolated NP is shown in the Supplemental Material [42].Finally, the morphology of the NPs on silica was characterized using an FEI Nova NanoSEM 650 at an acceleration voltage of 5 kV to record the secondary electron images of the Cu NPs on the substrates.

### (c) Contact angle measurement

The CA measurements were performed using a Dataphysics OCA25 system. An automated syringe dropped 2 μL droplets of pure water (MilliQ) on the sample, where a camera recorded the pictures over several seconds. Immediately after the testing droplet was loaded onto the measured surface, the camera started to record the meniscus for 10s and made the process as a video. Then the measured SCA was derived after "2s" from the video, because the SCA is more "stable" after 2s of the loading of the testing droplet [51]. The injecting/withdrawing speeds for advancing/receding contact angle measurements were both 0.2 μL/s, allowing the movement of the TPL steadily and smoothly. The drop shape was analyzed based on the form of an ideal sessile drop, for which the surface curvature results only from the force equilibrium between surface tension and weight of the liquid drop. The values of the contact angle were obtained *via* a fit using the Young-Laplace (YL) equation based on the shape analysis of a complete drop, and also compared to the results obtained from the geometrical CA analysis. For every sample, the CA measurements were repeated for several drops on different sample areas. In addition, the temperature was 20 °C, while the relative humidity was around 50%.

**Figure Captions**

**Figure 1** Atomic force microscopy (AFM) and wettability studies of the PMMA surfaces with various exposure times to high-pressure $CO_2$ treatments (58 bar). (a) AFM image of the PMMA without high-pressure $CO_2$ treatment with scan area of $3 \times 3$ μm$^2$. (b) SCA of the PMMA without high-pressure $CO_2$ treatment. (c)-(f) Height distributions of PMMA samples with 0 min (no treatment), 5min, 30 min, and 90 min high-pressure $CO_2$ treatments, respectively, derived from AFM topography measurements with scan area of $3 \times 3$ μm$^2$.

**Figure 2** Morphology and wetting studies of the high (64%)/low (12%)-$C_N$ samples. (a) AFM image of the high (64%)-$C_N$ sample with scan area of $3 \times 3$ μm$^2$.(b) TEM image of the Cu NPs deposited on a TEM grid simultaneously with the NPs deposited on the PMMA films including also the SCA (bottom right corner) for the high (64%)-$C_N$ Cu NPs/PMMA sample. (c) TEM image of Cu NPs deposited on a TEM grid simultaneously with the NPs deposited on the PMMM films with the SCA indicated (in the bottom right corner) of the low (12%)-$C_N$ Cu NPs/PMMA sample. (d) Distribution of the diameter of NPs measured in Fig. 2(c). (e), (g), (i) AFM images (scan area: $3 \times 3$ μm$^2$) of low (12%)-$C_N$ Cu NPs/PMMA samples with 0 min (no treatment), 30 min, and 90 min high-pressure $CO_2$ treatments, respectively. (f), (h), (j) Apparent height distributions of the monodispersed NPs measured in Figs. 2(e), (g), (i), respectively.

**Figure 3** Morphology studies of the "intermediate" (37%)-$C_N$Cu NPs/PMMA samples. (a), (c), (e), (g) AFM images (scan area: $1 \times 1$ μm$^2$) of the Cu NPs/PMMA samples with 0 min (no treatment), 5 min, 30 min, and 90 min high-pressure $CO_2$ treatments, respectively. (b), (d), (f),

(h)Apparent height distributions of monodispersed NPs measured in Figs. 3(a), (c), (e), (g), respectively.

**Figure 4** Wetting studies of the "intermediate" (37%)-$C_N$ Cu NPs/PMMA samples with various medians of $H_N$, and the Cu NPs/PMMA samples with various $C_N$ before high-pressure treatments. (a) SCAs and CAH of the "intermediate" (37%)-$C_N$ samples with various medians of $H_N$, where the square (blue) spots refer to the SCAs, and the circle (red) ones refer to the CAH. (b) ACAs and RCAs of the "intermediate" (37%)-$C_N$ samples with various medians of $H_N$, where the green bars refer to the ACAs and the purple bars to the RCAs, and the dashed areas in the green bars refer to the CAH of the samples.(c) SCAs and CAH of the samples with various $C_N$ coverages, where the square (blue) spots refer to the SCAs and the circle (red) ones to the CAH. (d) ACAs and RCAs of the samples, where the green bars refer to the ACAs and the purples ones to the RCAs, as well as the dashed areas in the green bars refer to the CAH.

**Figure 5** Schematics for evaluation of nanomeniscus on the wettability of the Cu NP decorated PMMA films. (a) The configuration used to calculate contact angle with Eq. (4): $R$ is the radius of the NP, $r_1$ is the negative radius of curvature, $r_2$ (not shown) is the positive radius of curvature in the orthogonal direction to the plane, $h_1$ is the thickness of wetting film on the substrate, $h_2$ is the thickness of wetting film on the NPs. (b) The schematic of partial NP engulfment into the PMMA film: $H$ is the depth of the Cu NP submerged into PMMA, $H_N$ is the apparent height of the NP, and $\alpha$ is the angle between the baseline of PMMA and the tangent of the NP from the sectional view.

**Figure 6** Relationships between the wettability and morphology of Cu NP decorated PMMA surfaces. (a) The relationship between $\Delta(\cos\theta_N)$ and $H_N$ fitted by Eq. (6). (b) The relationship between SCA and $H_N$, showing the experimental data (red circles), the linear fit (red dashed line), and the function that corresponds to Eq. (1) with $\Delta(\cos\theta_N)$ given by Eq. (6) (blue solid line). (c) The relationship between SCA and $C_N$, showing the experimental data (red circles), the best fit according to Eq. (8) (red dashed line), and the best fit by introducing a factor $\kappa$ (which is positively correlated with the average number of the NP layer on the substrate) into Eq. (8) (blue solid line).

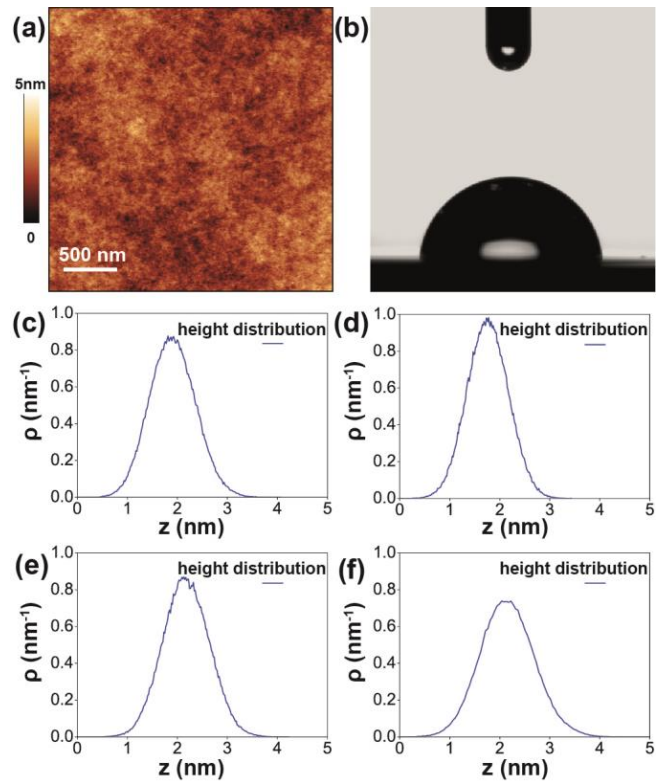

**Figure 1**

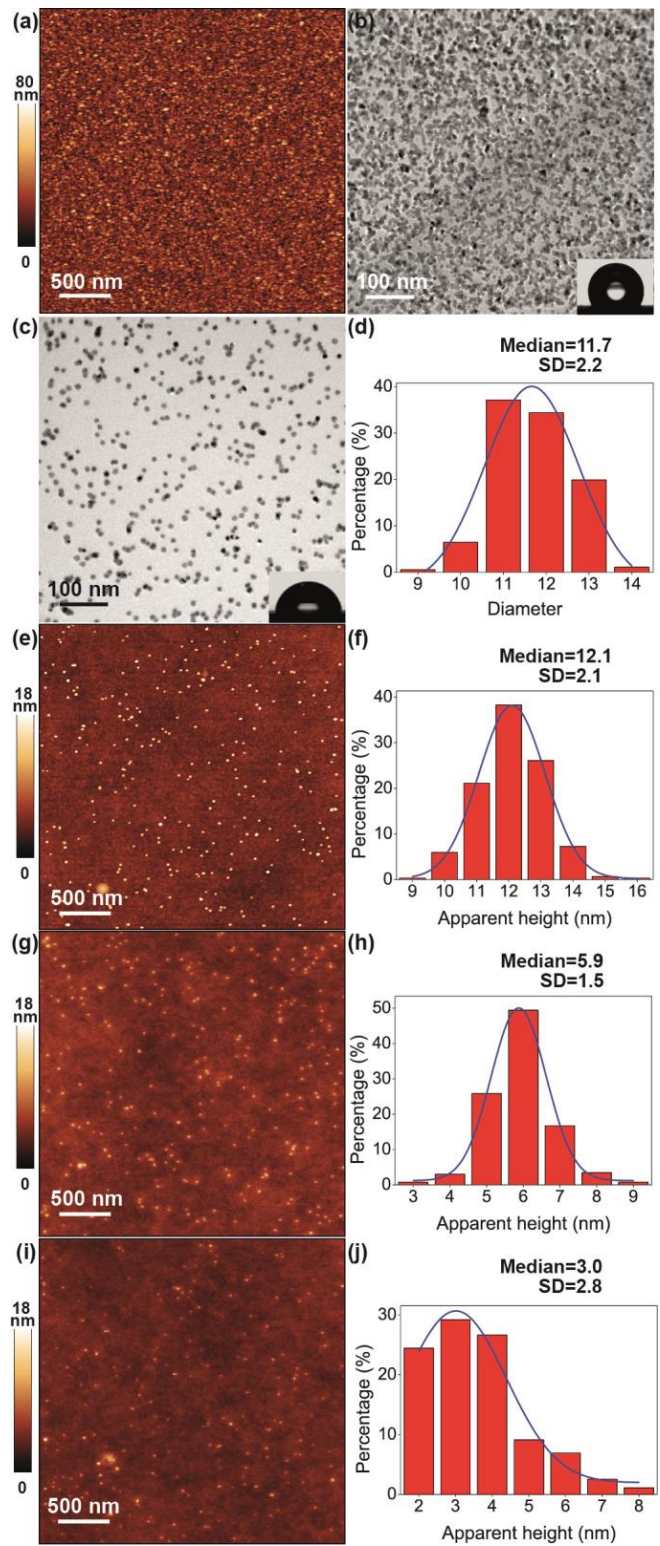

**Figure 2**

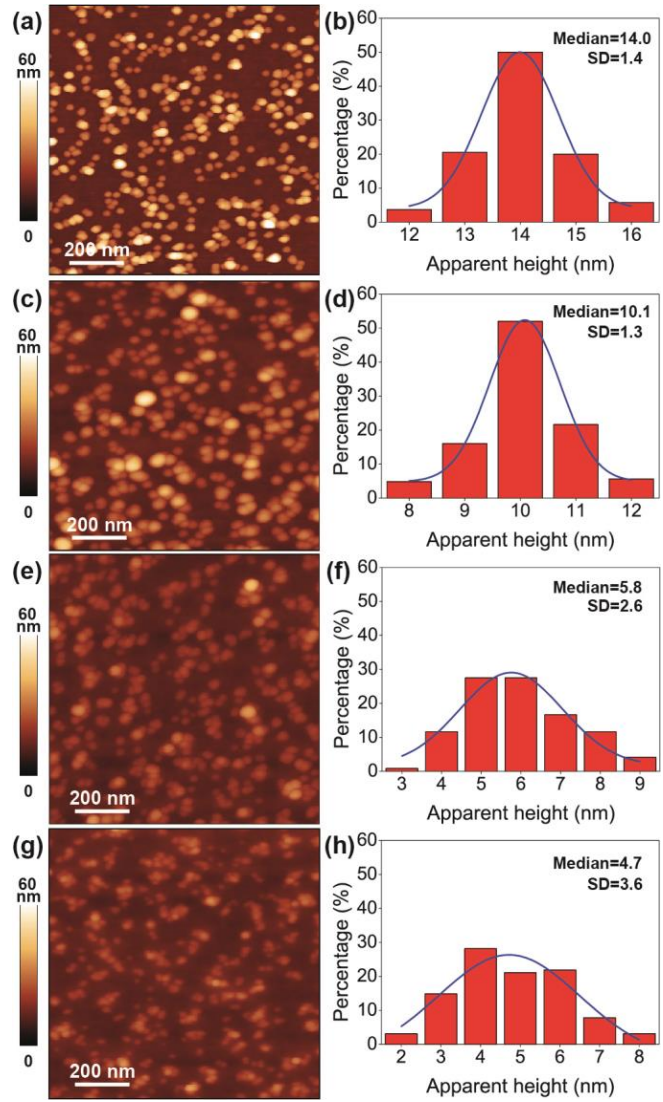

**Figure 3**

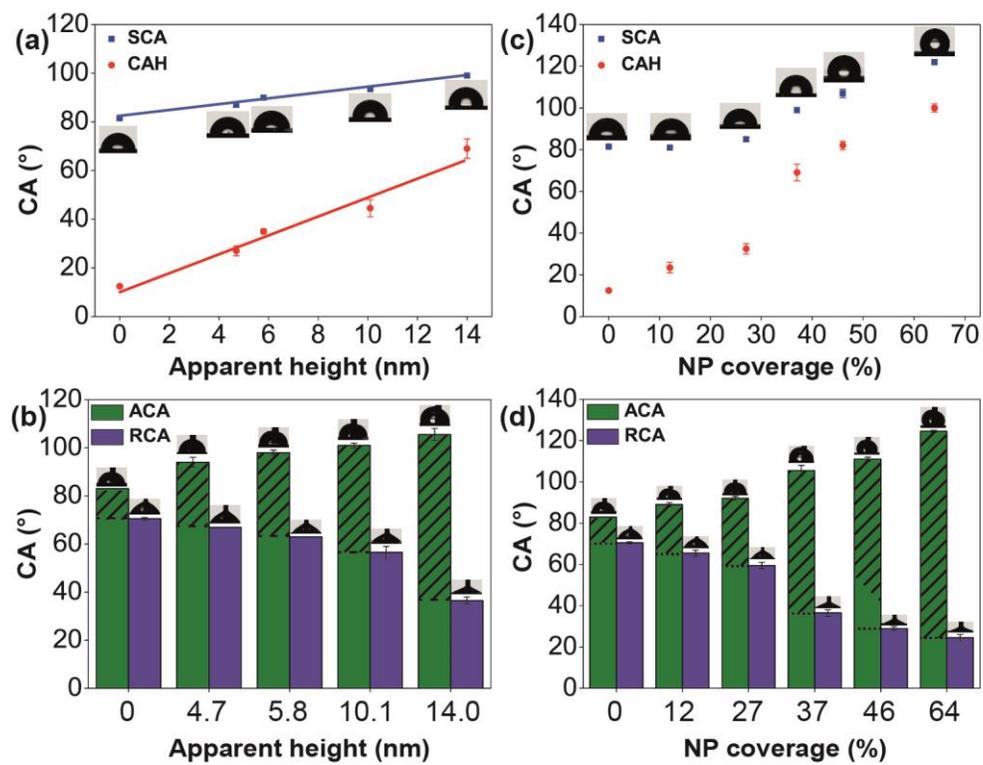

**Figure 4**

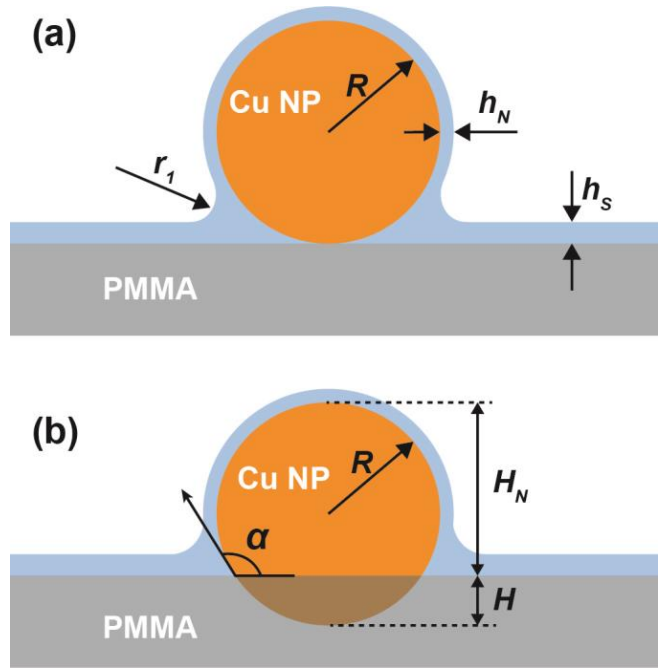

**Figure 5**

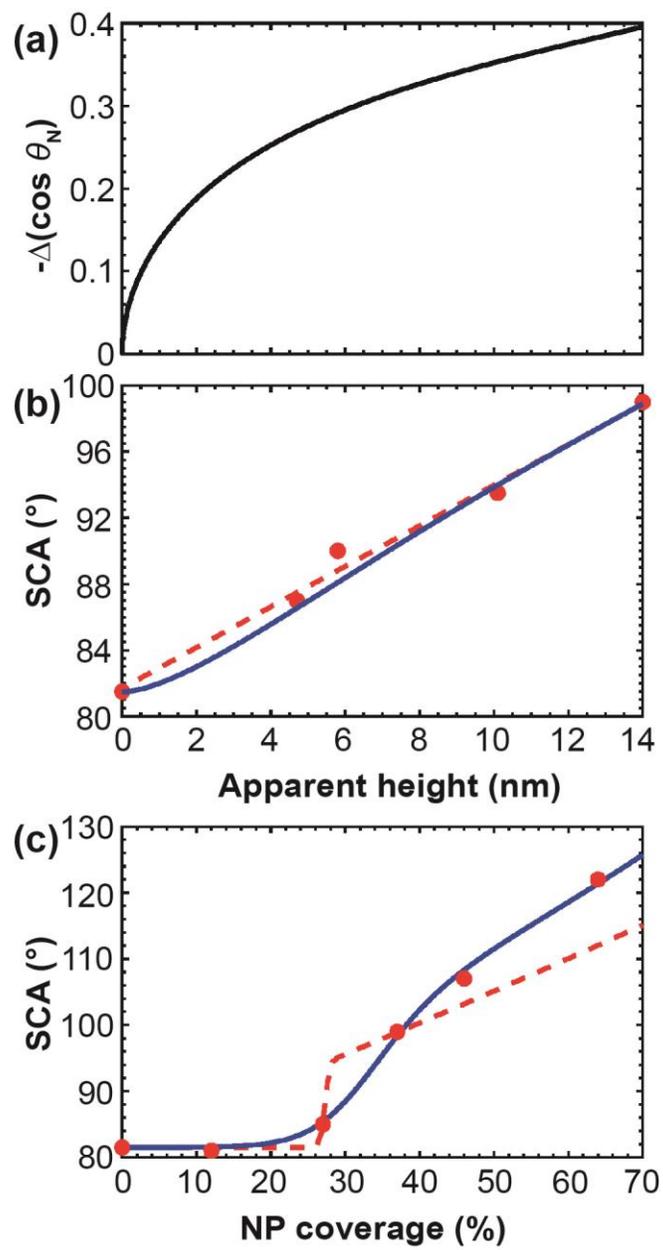

**Figure 6**

# Supplemental Material

**Tunable Wettability of Polymer Films by Partial Engulfment of Nanoparticles**

Weiteng Guo[1], Chongnan Ye[1], Gert H. ten Brink[1], Katja Loos[1], Vitaly B. Svetovoy[2], and George Palasantzas[1,*]

[1]Zernike Institute for Advanced Materials, University of Groningen, Nijenborgh 4, 9747 AG Groningen, The Netherlands

[2]A. N. Frumkin Institute of Physical Chemistry and Electrochemistry, Russian Academy of Sciences, Leninsky prospect 31 bld. 4, 119071 Moscow, Russia

# I. Effect of airborne hydrocarbons on surface wettability

So far, it has been proved that airborne hydrocarbons contaminants could have a significant effect on both flat surfaces as well as NP decorated surfaces [1]. Therefore, after each preparation, all the samples had been kept in air (Fig. S1) for more than one week, which is long enough for the as-prepared surfaces to adsorb enough hydrocarbons before the wetting measurements. In this way, the interference of the effect of airborne hydrocarbons on the wettability is excluded, ensuring the comparability of the measured wettability. In addition, all the samples also measured one month kept in air after the first measurements. And the data derived from the repeated measurements did not show any significant difference from the ones obtained from the first measurements. The latter indicates the successful exclusion of any interference effects from the airborne hydrocarbons.

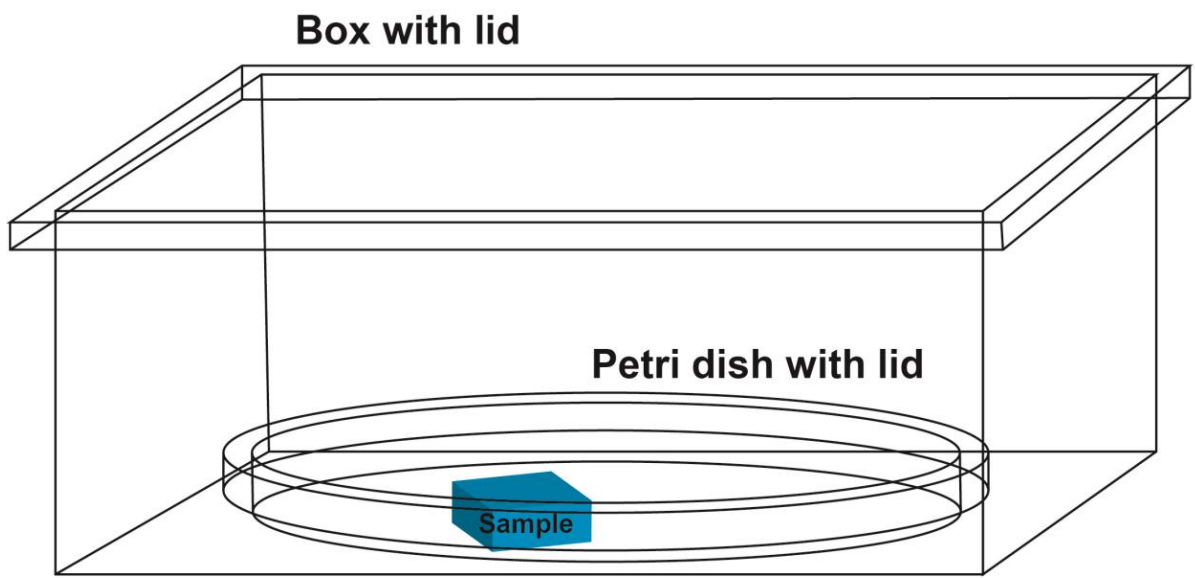

**FigureS1** Schematic of the containers for keeping the Cu NPs/PMMA samples.

As it is shown in Fig.S1, the Cu NPs/PMMA samples were kept in a petri dish with lid, and the dish was kept in a box with lid. Then the box was kept in a closet (which is clean and clear inside) in the laboratory. On one hand, this storage condition allow the sample to keep contact with ambient air and adsorb airborne hydrocarbons, on the other hand, this condition protect the samples from other contaminants, *e.g.*, dust and other particles in the open air.

## II. Measurement of the apparent height of an individual nanoparticle (NP)

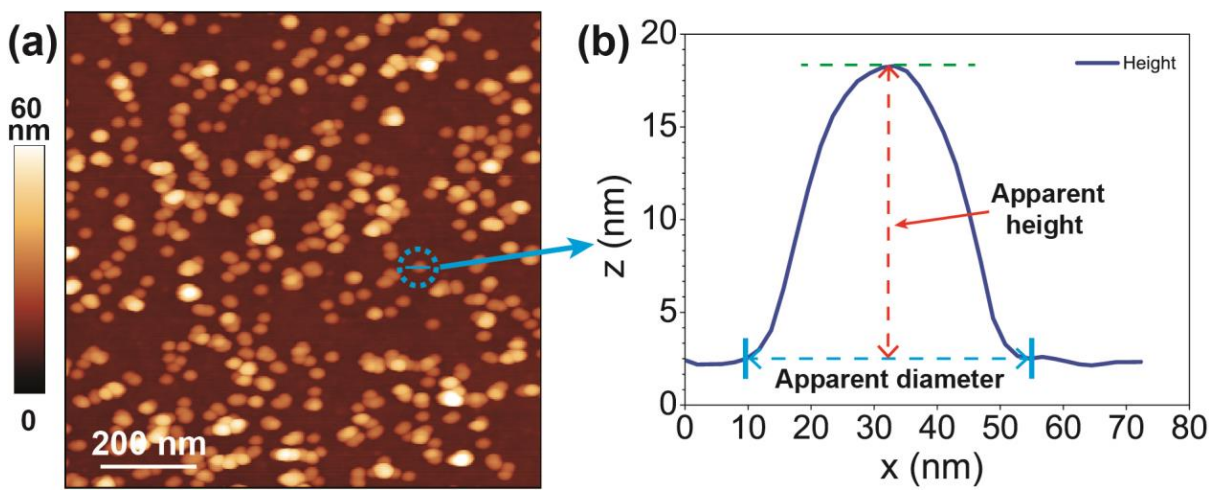

**Figure S2** Atomic force microscopy (AFM) topography measurements to obtain the apparent height of an individual isolated NP. (a) AFM image of an "intermediate" (37%)- $C_N$ Cu NPs/PMMA sample without high-pressure $CO_2$ treatment. (This is the same image as Fig. 4(a) in the manuscript) (b) The height profile extracted from a line scan (shown in the dashed circular area in Fig. S2(a)) through the center of an isolated NP.

A typical AFM measurement of an individual NP is shown in Fig. S2, where a hill-like profile was obtained as it is shown in Fig. S2(b). The distance (red dashed line in Fig. S2(b)) between the foot (blue dashed line in Fig. S2(b)) and the peak (green dashed line in Fig. S2(b)) of the "hill" was defined as the apparent height of the measured NP. In fact, the measured apparent height of the NP is 15nm. Additionally, the apparent diameter was defined as the distance (blue dashed line in Fig. S2(b)) between the two feet (two blue bars in Fig. S2(b)) of the "hill". As a result, the diameter of the NP was measured to be 45 nm. Comparing with the average diameter measured by TEM (~13.2 nm shown in the manuscript), it becomes clear that the lateral resolution of the AFM image is not as good as the vertical resolution.

### III. Determination of optimum NP coverage

As it emerges from the results in Fig. 2, a satisfactory Cu NPs/PMMA sample for extensive study should have a proper $C_N$ coverage to offer two main properties: One is having an obvious increase in SCA in comparison with the bare PMMA films (marked as Condition I), which can be achieved for higher $C_N$ samples, the other one is allowing the measurement of the NP height through AFM (marked as Condition II), which requires lower $C_N$ samples. The two mutually contradictory requirements call for a sample with an "intermediate" $C_N$. Additionally, there is still a supplementary requirement for the $C_N$ that the PMMA films should be covered mainly by a monolayer of NPs (marked as Condition SI), which will offer convenience for the theoretical analysis. Besides the high and low $C_N$ samples shown in Fig. 2, three more trials were conducted to determine the Cu NPs/PMMA samples with the proper $C_N$ for the subsequent research. Afterwards, two Cu NPs/PMMA samples with $C_N \approx 46\%$ (termed as "higher" (46%)-$C_N$ sample) and $C_N \approx 27\%$ (termed as the "lower" (27%)-$C_N$ sample) were prepared but could not meet all the requirements, indicating the "proper" $C_N$ coverage of NPs might be somewhere in the range of 27%~46%. Figs. S2a, b were derived from the same Cu NPs/PMMA sample with $C_N \approx 46\%$ (termed as "higher" (46%)-$C_N$ sample). The SCA of the sample was 107±2° (see Fig. S2a), showing a significant increase to address the requirement of Condition I. Nevertheless, the sample did not meet Condition II because the Cu NPs shown in the AFM image (Fig. S2b) were too dense to be measured as isolated NPs. Additionally, there were areas with double-layer NP as it is shown in the TEM image (Fig. S2a), which did not meet Condition SI. Figs. S2c, d were derived from the same Cu NPs/PMMA sample with $C_N \approx 27\%$ (termed as the "lower" (27%)-$C_N$ sample). And most of the Cu NPs in the AFM image (Fig. S2d) were isolated NPs, which enabled reliable AFM analysis meeting Condition II. Besides, the "lower" (27%)-$C_N$ sample also met Condition SI because the majority of the area shown in Fig. S2c was covered with a monolayer of NPs. However, the "lower" (27%)-$C_N$ sample did not meet Condition I because the SCA of the sample (85±1° shown in Fig. S2c) did not represent any significant increase compared with the bare PMMA films (SCA≈81.5±0.5°). It could be concluded from the results in Figs. S2a-d that neither the "higher" (46%)-$C_N$ sample and nor the "lower" (27%)-$C_N$ sample met all the requirements for any subsequent further study. Nevertheless, the results of these two samples indicated that the "proper" $C_N$ coverage of NPs might be somewhere in the range of 27%~46%.

Subsequently, "intermediate"-$C_N$ Cu NPs/PMMA samples were made and characterized. To begin with, the $C_N$ of the "intermediate"-$C_N$ sample was measured as 37% (termed as "intermediate" (37%)-$C_N$ sample) in Fig. S3(e), and the majority of the area was covered with a monolayer of NPs meeting the Condition SI. Moreover, the SCAs of the "intermediate" (37%)-$C_N$ Cu NPs/PMMA samples were measured around 99°, which is significantly different from the bare PMMA films (81.5±0.5°), meeting Condition I. Finally, most of the Cu NPs in the AFM image (Fig. 3(a)) were isolated, which enabled reliable AFM analysis, meeting Condition II. In short, the "intermediate" (37%)-$C_N$ Cu NPs/PMMA samples met all the conditions and offered the "proper" samples for our subsequent extensive study.

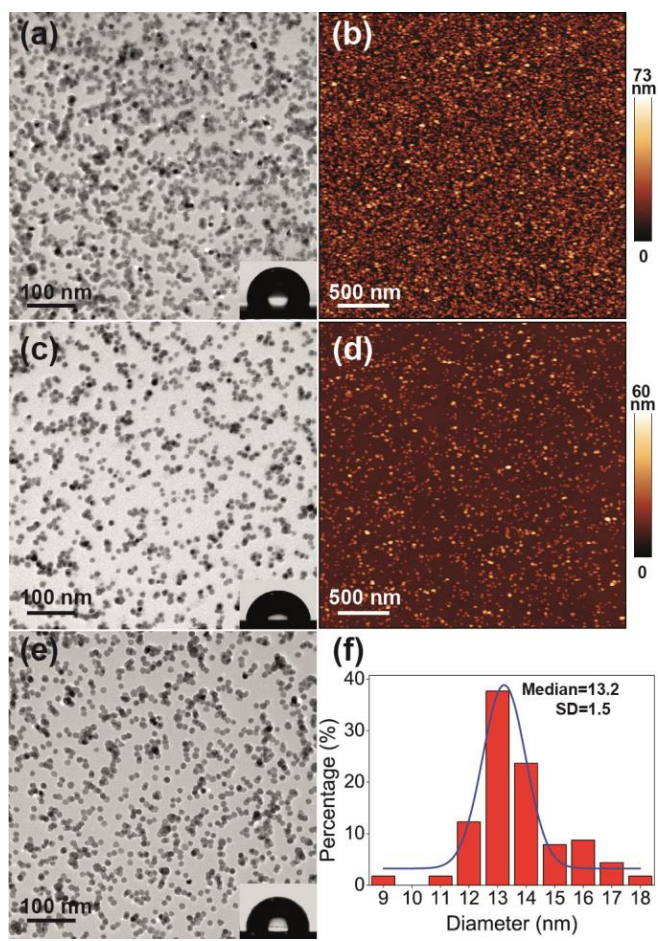

**Figure S3** Morphology and wetting studies of the Cu NPs/PMMA samples with various $C_N$ coverages. (a) TEM image of Cu NPs deposited on a TEM grid simultaneously with the NPs deposited on the PMMM films together with the SCA (bottom right corner) of the "higher" (46%)-$C_N$ Cu NPs/PMMA sample. (b) AFM image of the "higher" (46%)-$C_N$ Cu NPs/PMMA sample with a scan area of 3 × 3 μm². (c) TEM image of Cu NPs deposited on a TEM grid simultaneously with the NPs deposited on the PMMM films together with the SCA (bottom right corner) of the "lower" (27%)-$C_N$ Cu NPs/PMMA sample. (d) AFM image of the "lower" (27%)-$C_N$ Cu NPs/PMMA sample with a scan area of 3 × 3 μm².(e) TEM image of Cu NPs deposited on a TEM grid simultaneously with the NPs deposited on the PMMM films together with the SCA (bottom right corner) of the "intermediate" (37%)- $C_N$ Cu NPs/PMMA sample. (f) Distribution of the diameter of NPs measured in Fig. S3(e).

## IV. AFM images for improving the statistics

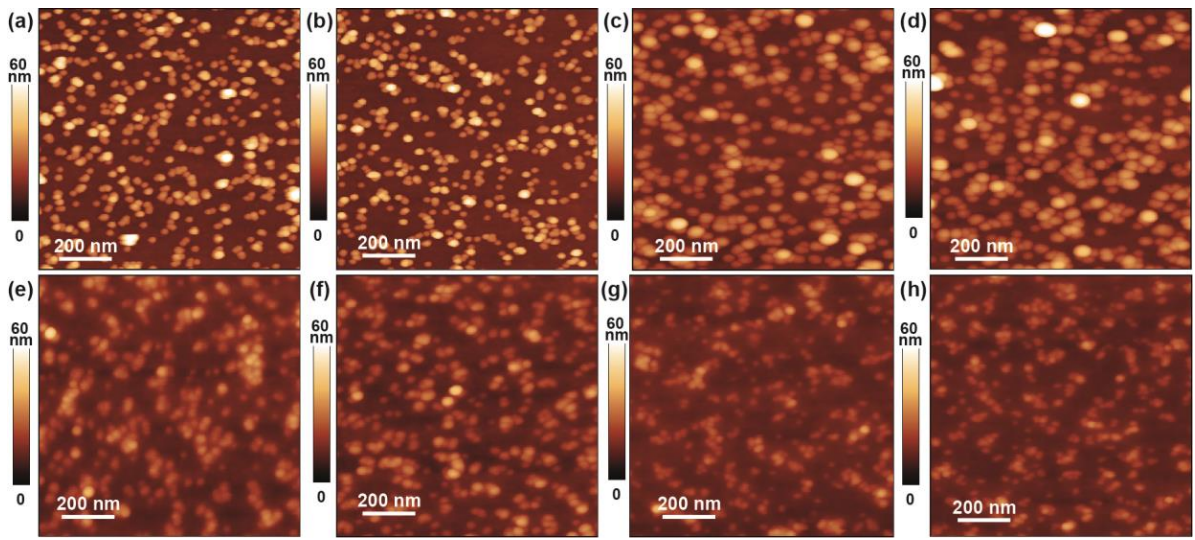

**Figure S4** Additional AFM images obtained from the same samples used in Fig.4 in the manuscript for improving the statistics. (a), (b) AFM images of the pristine Cu NPs/PMMA sample. (c), (d) AFM images of the Cu NPs/PMMA sample with 5 min high-pressure $CO_2$ treatment. (e), (f) AFM images of the Cu NPs/PMMA sample with 30 min high-pressure $CO_2$ treatment. (g), (h) AFM images of the Cu NPs/PMMA sample with 90 min high-pressure $CO_2$ treatment.

## V. Derivation of Eq. (2)

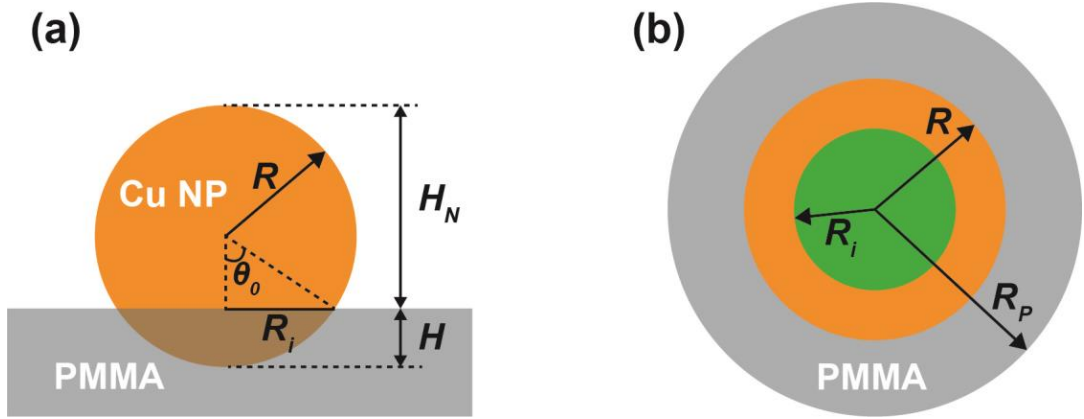

**Figure S5** Schematics of the calculations of the ratios of the true area of the solid surface to the apparent area of Cu NPs/PMMA samples with NP engulfment. (a) Sectional view. (b) Top view.

Imagining that the whole PMMA substrate was homogeneously divided into $n$ ($n$ is the number of the decorating NPs) circular parts with radii $R_P$, and each part has an NP standing on it. Then we take each part as a "unit" (the shape of the "unit" doesnot affect the result, but the circular shape would simplify the process of the calculations). Since every "unit" share the same property, the ratios of the true area of the solid surface to the apparent area of Cu NPs/PMMA samples, $f_S$ for the substrate and $f_N$ for the NPs, are identical to the ones of the "unit". Here we assume the shape of the PMMA surface of the "unit" as a circle, then the calculation was simplified, as it is shown in Fig. S5. A spherical NP of radius $R$ was submerged into the substrate to a depth $H$, and the circular interface (interfacial area) of the NP and the substrate with radius $R_i$ is more clearly shown in Fig. S5(b), and it is marked with green color. Subsequently, the ratio of the apparent area of the substrate, $f_S$, is given by

$$f_S = \frac{A_P - A_i}{A_P} = 1 - \frac{A_i}{A_P} = 1 - \frac{\pi R_i^2}{\pi R_P^2}, \tag{S1}$$

where $A_P$ is the area of the PMMA surface prior to NP deposition, and $A_i$ is the area of the interface of the NP and the PMMA. As it is shown in Fig. S5(a), $R_i^2 = R^2 - (R-H)^2$. Then Eq. (S1) can be rewritten as

$$f_S = 1 - \frac{R^2 - (R-H)^2}{R_P^2}. \tag{S2}$$

Since the NP coverage ($C_N$) is defined as $C_N = \pi R^2 / \pi R_P^2$, Eq. (S2) is rewritten as

$$f_S = 1 - \frac{R^2 - (R-H)^2}{R^2/C_N} = 1 - C_N \frac{H(2R-H)}{R^2}. \tag{S3}$$

As it is shown in Fig. S5(a), the ratio of the apparent area of the NP, $f_N$, is given by

$$f_N = \frac{A_N}{A_P}, \tag{S4}$$

where $A_N$ is the surface area of the apparent part of the NP that can be calculated by

$$A_N = \int_0^{2\pi} d\varphi \int_{\theta_0}^{\pi} R^2 \sin\theta \, d\theta, \tag{S5}$$

with $\cos\theta_0 = (R-H)/R$. Integration of Eq. (S5) yields:

$$A_N = 2\pi R(2R - H), \tag{S6}$$

And Eq. (S4) can be rewritten as

$$f_N = \frac{2\pi R(2R-H)}{\pi R^2/C_N} = C_N \frac{2(2R-H)}{R}. \tag{S7}$$

Therefore, Eq. (2) in the manuscript is derived.

## VI. Movability of NPs on PMMA

It has been proved that NPs on flat substrates can move if a lateral force was applied to the NPs [2]. Therefore, scanning electron microscopy (SEM) and AFM measurements were conducted to the residual parts of the CA measured areas of the samples to check the movability of NPs on flat substrates under CA measurement, and discuss the effect of this movability to our results. However, both methods failed to characterize the samples for the following reasons: For SEM measurement, a gold layer were deposited to the samples through magnetron sputtering because PMMA is not conductive. When a relatively thin layer of gold (5-10 nm) was deposited, the area scanned by the electrons would crack quickly because high magnification was applied to observe the NPs and the current of the electron beam was too concentrated and able to break the structure of the PMMA. Then the thickness of the deposited gold layer was enhanced to 20 nm to protect the sample during SEM measurement, the cracking phenomena was prevented but the morphology features of NPs were "buried" by the thicker gold layer. Subsequently, AFM measurements were used to observe the areas of the wetted areas left by a testing droplet. However, the wetted areas were more "blurred" than the unwetted areas, and we attributed this phenomena to some nanoscale contaminants which were formed during/after contact angle measurement. The nanoscale contaminants were observed by SEM measurement in the Fig. S6, and the contaminants "blurred" the AFM image because the lateral resolution of AFM was limited by the "tip-effect" preventing to distinguish NPs and the "new-born" nanoscale contaminants. Therefore, we could not derive any conclusion if the NPs could move above PMMA by the AFM measurements.

Therefore, the characterization was conducted indirectly, and Cu NPs/silica samples were prepared because silica wafers are smooth (which have similar morphology as the drop casted

PMMA films in this work) and conductive, which are visible in SEM. Subsequently, the wetted areas left by static and dynamic CA measurements were observed through SEM and the results are shown in the Fig. S6. In Figs. S6(a)-(d), each SEM image was "divided" into two parts by a line in the middle, which referred to the boundary of the wetted area left by the CA measurement. The left part referred to the wetted area, while the right part referred to the unwetted area. The NP sliding phenomena were not obvious during the static CA measurement shown by Figs. S6(a), (b). Nevertheless, an obvious NP concentration was observed at the edge of the wetted area left by a dynamic CA measurement (see Figs. S6(c), (d)), indicating that NP sliding took place during dynamic CA measurements. Furthermore, it is known that the movability of NPs on substrate is closely related to the adhesion force between the NPs and substrate [3]. The substrate hardness is the most significant factor in the NPs adhesion to the substrate based on previous studies. Comparing the hardness of the substrates we used, PMMA, (70-100 MPa) and $SiO_2$ (5-10 GPa), the hardness of PMMA is much lower than the one of $SiO_2$, indicating that the Cu NP adhesion force to PMMA substrate is much higher than the NP adhesion force the $SiO_2$ substrate, which makes the Cu NPs much more difficult to slide on PMMA during the CA measurements. Therefore, it is reasonable to presume that NP movement on PMMA was not obvious during static CA measurement, and the NPs might be "pushed" by the TPL of the droplet during dynamic CA measurements. However, the NP sliding phenomena on PMMA would be less evident than the one on silica (which we actually observed in SEM). Nevertheless, after various time-span high-pressure $CO_2$ treatments, the NPs were more or less embedded into PMMA, which made the NPs hard to move during wetting measurements. This is also an advantage of high-pressure $CO_2$ induced NP engulfment, which can stabilize the NPs on PMMA and eliminate the NP-sliding factor from the experiments. Therefore, even if we exclude the data point belonging to the Cu NPs/PMMA sample prior to $CO_2$ treatment, we could still obtain similar tendency of the relationship between wettability and the apparent height of NPs. Besides, in Fig. 4(a), we can also extrapolate the data point belonging to the Cu NPs/PMMA sample prior to $CO_2$ treatment through the other data points. The negligible difference between the experimental result and the extrapolated data indicates that the NP sliding phenomena, if any, does not have significant influence on the conclusions we have derived.

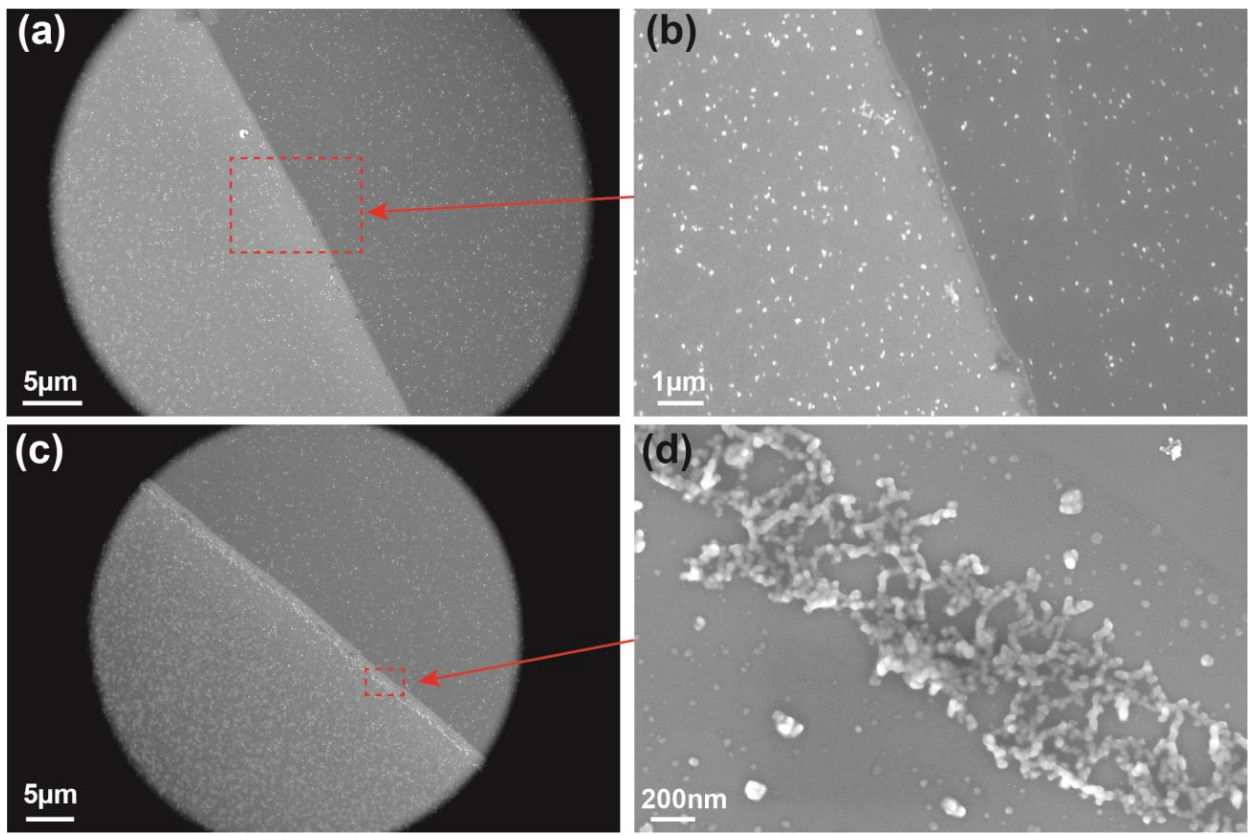

**Figure S6** SEM studies of the boundaries of the wetted areas left by CA measurements. (a) Low-magnification SEM image of the boundary of a wetted area left by a static CA measurement. (b) Locally magnified SEM image of the area referring to the area in the red box in Fig.S6(a). (c) Low-magnification SEM image of the boundary of a wetted area left by a dynamic CA

measurement. (d) Locally magnified SEM image of the area referring to the area in the red box in Fig. S6(c).

## VII. Possible polymer layer formed on NPs during NP engulfment

It has been reported that NPs (20 nm in size gold) would be rapidly covered by a thin polymer wetting layer with a thickness of 1.3-1.8 nm upon annealing above the glass transition temperature of polystyrene (PS) [4]. Therefore, it is reasonable to assume that a similar PMMA film might also form on Cu NPs during the high-pressure $CO_2$ treatment of Cu NPs/PMMA samples in this work. Subsequently, it is necessary to discuss the possible effect of this thin film on this work. The polymer layer covering the NPs was very thin and uniform, so that the shape of the NPs did not change significantly [4]. On the other hand, the wettability of Cu (79°) and PMMA (81.5±0.5°) are similar. Therefore, the cover of a PMMA film would not make a significant change to the wettability of Cu NPs. Besides, in Fig. 4(a), we can also extrapolate the data point belonging to the Cu NPs/PMMA sample prior to $CO_2$ treatment through the other data points. The negligible difference between the experimental result and the extrapolated data indicates that the thin PMMA film on Cu NPs, if any, does not have significant influence on the conclusions we have derived.